\documentclass{article}

\usepackage{indentfirst} %\linespread{1.3}
\usepackage{amsmath}
\usepackage{epsfig}
\usepackage{amssymb}
\usepackage{graphicx}
\usepackage{color}
\usepackage{caption}
\usepackage{subfigure}
\usepackage{amsmath}
\usepackage{epsfig}
\usepackage{amssymb}
\usepackage{color}
\usepackage{algorithm}
\usepackage{algorithmic}
\usepackage{multirow} %multirow for format of table
\usepackage{amsmath}
\usepackage{xcolor}
\usepackage{graphicx}          % Include this line if your
\usepackage{booktabs}
\usepackage{array}
\usepackage{epstopdf}
\usepackage{color}
\usepackage[amsmath,thmmarks]{ntheorem}
\usepackage{arxiv}
\usepackage{graphicx,natbib,amssymb,amsmath,bm}
\usepackage[utf8]{inputenc} % allow utf-8 input
\usepackage[T1]{fontenc}    % use 8-bit T1 fonts
\usepackage{hyperref}       % hyperlinks
\usepackage{url}            % simple URL typesetting
\usepackage{booktabs}       % professional-quality tables
\usepackage{amsfonts}       % blackboard math symbols
\usepackage{nicefrac}       % compact symbols for 1/2, etc.
\usepackage{microtype}      % microtypography
\usepackage{lipsum}		% Can be removed after putting your text content
\usepackage{graphicx}
\usepackage{natbib}
\usepackage{algorithm}
\usepackage{algorithmic}
\usepackage{epstopdf}
\usepackage{doi}
\usepackage{subeqnarray}
\usepackage[amsmath,thmmarks]{ntheorem}

\title{A distributionally robust optimization approach to two-sided chance constrained stochastic model predictive control  with unknown noise distribution}

\author{{\hspace{1mm}Yuan~Tan} \\
		The School of Automation\\
	Southeast University\\
	Nanjing, China \\
    \texttt{tan\_yuan@aliyun.com} \\
	\And
	{\hspace{1mm}Jun~Yang} \thanks{Corresponding author}\\
	DThe School of Automation\\
	Southeast University\\
	Nanjing, China \\
	\texttt{j.yang3@lboro.ac.uk} \\
		\And
	{\hspace{1mm}Wen-Hua~Chen} \\
	Department of Aeronautical and Automotive Engineering\\
	Loughborough University\\
	Loughborough, LE11 3TU, UK \\
	\texttt{W.Chen@lboro.ac.uk} \\
	\And
		{\hspace{1mm}Shihua~Li} \\
		The School of Automation\\
	Southeast University\\
	Nanjing, China \\
	\texttt{lsh@seu.edu.cn} \\
	}

\renewcommand{\shorttitle}{\textit{arXiv} Template}

\hypersetup{
pdftitle={A distributionally robust optimization approach to two-sided chance constrained stochastic model predictive control  with unknown noise distributio},
pdfauthor={Yuan~Tan, Jun Yang, Wen-Hua Chen, Shihua Li},
pdfkeywords={Distributionally robust optimization; Stochastic model predictive control; Two-sided chance constraints; Second-order cone.},
}

\begin{document}
\maketitle

\begin{abstract}
In this work, we  propose a distributionally robust stochastic model predictive control (DR-SMPC) algorithm to address the problem of two-sided chance constrained discrete-time linear system corrupted by additive noise. The prevalent mechanism to cope with two-sided chance constraints is the so-called risk allocation approach, which conservatively approximates the two-sided chance constraints with two single chance constraints by applying the Boole's inequality.
In this proposed DR-SMPC framework, an exact tractable second-order cone (SOC) approach is adopted to abstract the two-sided chance constraints by considering the first and second moments of the noise. The proposed DR-SMPC algorithm is able to guarantee that the worst-case probability of violating both the upper and lower limits of safety constraints is within the pre-specified maximum probability (PsMP). By flexibly adjusting this PsMP, the feasible region of the initial states can be increased for the SMPC problem. The recursive feasibility and convergence of the proposed DR-SMPC are established rigorously by introducing binary initialization strategy of nominal state.  Simulation studies of two practical cases  are conducted to demonstrate the effectiveness of the proposed DR-SMPC algorithm.
\end{abstract}

% keywords can be removed

\section{Introduction}
Conventional control methods (e.g., linear quadratic regulator and dynamic programming) have been extensively investigated to solve the problem of a discrete-time linear system. However, these control techniques are not able to deal with constraints on the system state or the control input which becomes more and more important due to the intensifying pace towards era of safe autonomy.  To address this challenge, as a promising and efficient solution, model predictive control (MPC) has been proposed to solve a finite horizon, constrained optimization control problem at each sampling time and to determine a finite sequence of control actions. MPC has attracted considerable attention to both industry and academia over the last couple of decades \citep{Rawlings}.

In almost all practical applications, the behavior of the system is falsified by various uncertainties, e.g.,
unknown parameters, external disturbances, and process noises.  In the presence of uncertainties, the controller may fail to either guarantee safe operation or meet quality specifications. If the bound of the uncertainties can be quantified or know a priori, deterministic robust model predictive control (RMPC) approaches have been developed to address these intractable uncertainties \citep{2005Robust,2007Robustness}. It should be highlighted that in the context of RMPC, the robust constraints satisfaction, recursive feasibility and stability are all established in a conservative manner by solving a min-max optimization problem $-$ that is minimizing the cost but considering the possible maximum impacts (worst-case) of uncertainties. Furthermore, the complicated inflation of the uncertainty quantification set throughout system dynamics, safe constraints, optimization and control loops may result in a very small feasible region of the initial states and even infeasible solutions of the optimization problem.

To tackle these drawbacks, stochastic model predictive control (SMPC) delivers a promising way to reduce the inherent conservativeness of RMPC by developing chance constraints that allow the constraints to be violated with a pre-specified maximum probability (PsMP) \citep{Mesbah2016Stochastic,2016Stochastic,2012Nonquadratic}. SMPC provides an adequate way to achieve trade-off between the closed-loop control performance and the constraints satisfaction \citep{2014Stochastic}. The potential of chance constraint quantification can be seen from many practical applications, e.g., concentrations on a chemical reactor control \citep{2016Stochastic} or comfort on robot obstacle avoidance \citep{S}. As a byproduct of chance constraints, the feasible set of the initial state can be enlarged under SMPC even without changing the prediction horizon.

 In recent years, there have been some achievements to address SMPC problem. Broadly speaking, SMPC approaches can be classified into two categories \citep{2016Stochastic}. The first category is called randomized approaches \citep{2012Randomized,2019Scenario,2014The,SHANG201924,Mark2020StochasticMW}.  The randomized approaches use samples/scenarios of the noise to approximate the SMPC problem.  Probabilistic closed-loop guarantees can be established by using scenario optimization tools \citep{2019Scenario}.  However, this needs a large amount of sampling data so that it consumes considerable computing time. Furthermore, it is generally quite hard to rigorously establish recursive feasibility and closed-loop stability of these approaches \citep{Mark2020StochasticMW}.

The second category is referred to as analytical approximation approaches \citep{2012Nonquadratic,2019Piecewise,2013A,2015An,HEWING2020109095}. In this setting of approaches, if a prior distributional information about the noise is known, the inverse of the cumulative probability function can be used to reformulate the chance constraints \citep{2012Nonquadratic,2019Piecewise}. However, it is usually quite difficult to know the true probability distribution of process noise in practice, which may result in  infeasibility or undesirable behavior of the system under SMPC. A milestone in this area \citep{2013A,2015An} is to use Cantelli's inequality to approximate the chance constraints by taking into account the mean and variance of noise. Despite the computing burden of the SMPC is greatly reduced, it can only deal with one-sided chance constraint.  However, in practical applications, two-sided constraints are pervasive, e.g., the acceleration of vehicles \citep{7733074} and the temperature change of the the building heating room \citep{SHANG201924} can be either positive or negative. Consequently, it is of great practical significance to investigate two-sided chance constrained SMPC problem.  One feasible manner to deal with two-sided chance constraints is to develop risk allocation mechanisms - that is decomposing the constraints into a set of one-sided constraints which conservatively approximate the original chance constraints.

The risk allocation mechanisms mainly follow two strategies below. The first strategy usually uses a uniform allocation strategy to obtain a fixed violation probability value for each one-sided constraints \citep{2007Convex}. This approach simplifies the optimization problem, however it may lead to significant conservatism in many situations due to the loss of active risk allocation. To address this limitation, the second strategy takes the risk allocation as the decision variables of the optimization problem \citep{2013Ao,article}. The risk allocation mechanisms brought by the Boole's inequality essentially narrow feasible set of states on SMPC scheme.

To overcome the conservativeness for two-sided chance constraints handling of existing SMPC approaches, a distributionally robust optimization (DRO) approach is proposed to solve the two-sided chance constrained SMPC problem. Distributionally robust chance constraints (DRC2s) have direct connection to the chance constraints incorporated in the classical paradigms of  SMPC  \citep{2019Distributionally}. DRC2s assume the actual distribution of noise belongs to an ambiguity set which contains all distributions with a predefined characterizations (e.g., first or second moments of noise). Given the aforementioned benefits, DRC2s are considered in this work, where a less conservative approximation approach for DRC2s is developed by using a second-order cone (SOC) reformulation. Consequently, this proposed distributionally robust SMPC (DR-SMPC) can be reformulate as a convex optimization problem. While assigning the value of the true measurement state to the nominal state may  lead to infeasible optimization problem, a binary initialization strategy is further developed to determine the nominal state, guaranteeing recursive feasibility and convergence for the proposed method.

The differences between this paper and other related DRO works (e.g., \citep{Mark2020StochasticMW,2016Distributionally,TY}) are summarized as follows. In \cite{Mark2020StochasticMW} and \cite{TY}, the DRC2s are first introduced into SMPC problem. Only one-sided chance constraint is reformulated into computable expression which are incapable to deal with two-sided chance constraints. In \cite{2016Distributionally},  a SOC programming reformulation of a general two-sided DRC2s is proposed to solve an optimal power flow problem, which is a normal optimization problem rather than in the context of a receding horizon optimization paradigm. While this paper proposes a new mechanism to address a two-sided distributionally robust chance constrained SMPC problem. Furthermore, specific form of terminal constraint for state is imposed, and the recursive feasibility and convergence are established. The main contributions are threefold, which are summarized below.
\begin{enumerate}
    \item The proposed approach delivers a tractable conic optimization solution to handle two-sided chance constraints by using  an ambiguity set the first and second moments of noise.
    \item A less conservative approach based on SOC is developed to abstract the DRC2s, which provides larger initial feasible set as compared to the existing risk allocation approaches.
    \item The recursive feasibility and convergence is rigorously established by introducing a binary initialization strategy to determine the nominal state.
\end{enumerate}

The rest of this paper is organized as follows.   We state the two-sided chance-constrained SMPC problem in Section II. Section III shows {\bf DR-SMPC-P1} can be converted into a convex SOC program. The recursive feasibility and convergence of the algorithm are  proved in Section IV.   Simulation results are reported in Section V to demonstrate the effectiveness of the proposed  DR-SMPC algorithm.  Finally, we end this paper by making some concluding remarks in Section VI.

\section{Problem Statement}
\subsection{Stochastic System}
We consider a discrete-time linear system with additive stochastic noise
\begin{equation}\label{T1}
x_{k+1}=A x_k+Bu_k+w_k
\end{equation}
where $k\in \mathbb{N}_0$ ($\mathbb{N}_0=\left\{0,1,2,\cdots\right\}$ is a set of nonnegative integer) is discrete-time index,  $x_k\in {\mathbb R}^{n_x}$ is the state, and $u_k\in {\mathbb R}^{n_u}$ is the input. $w_k\in {\mathbb R}^{n_x}$ is an unknown stochastic noise with known mean $\mu=0$ and covariance matrix $W\succeq 0$. The known pair $(A,B)$ is assumed to be stabilizable.
\newtheorem{assumption}{Assumption}

Given the state $x_k$ at time step $k$,  the predicted state is able to be updated according to
\begin{equation}\label{YT1}
x_{l+1|k}=A x_{l|k}+Bu_{l|k}+w_{l|k}, x_{0|k}=x_k,
\end{equation}
where $l|k$ ($l\in\mathbb{N}_{0}$) is the predicted $l$ steps ahead at time step $k$.
\subsection{Two-sided DRC2s}
The state and input are subject to chance constraints
\begin{align}
&{\rm Pr}_{[{\mathbb P}]}  \left[x_{l|k}\in \mathcal{X}|x_k\right] \geq 1-p_x,l\in \mathbb{N}_0;\label{T2}\\
&{\rm Pr}_{[{\mathbb P}]}  \left[u_{l|k}\in \mathcal{U} |x_k\right] \geq 1-p_u,l\in \mathbb{N}_0,\label{T3}
\end{align}
where ${\rm Pr}_{[{\mathbb P}]}\left[ \cdot \right] $ denotes the probability under the distribution ${\mathbb P}$ of uncertainties, $\mathcal{X}=\left\{x\in \mathbb{R}^{n_x}| H^Tx\leq h\right\}$  and $\mathcal{U}=\left\{u\in \mathbb{R}^{n_u}| F^Tu\leq f\right\}$ are convex sets containing the origin in its interior with constant matrices $H^T \in \mathbb{R}^{p \times n_x}$, $F^T \in \mathbb{R}^{q \times n_u}$ and  constant vectors $h\in \mathbb{R}^p$, $f\in \mathbb{R}^q$,   and $p_x, p_u\in(0,1]$ are the PsMP that the constraints $x_{l|k}\in \mathcal{X}$ and $u_{l|k}\in \mathcal{U}$ are allowed to be violated.

Chance constraints are also named as Value-at-Risk (VaR) constraints, which are in general non-convex feasible set in SMPC optimization problem.
In most SMPC approaches, the chance constraints are reformulated into deterministic feasible set through a common assumption that the probability distribution ${\mathbb P}$ of stochastic noise  $w_k$ is known. However, this does not match the noise in the real environment where the true distribution ${\mathbb P}$ is unknown.
Since this is an undesirable fact,  a distributionally robust version of (\ref{T2}) and (\ref{T3}) in  \cite{Mark2020StochasticMW,TY} is  introducing as
\begin{align}
&\inf_{{\mathbb P}\in{\mathcal P}} {\rm Pr}_{[{\mathbb P}]}  \left[x_{l|k}\in \mathcal{X}|x_k\right] \geq 1-p_x,l\in \mathbb{N}_0;\label{T5}\\
&\inf_{{\mathbb P}\in{\mathcal P}} {\rm Pr}_{[{\mathbb P}]}  \left[u_{l|k}\in \mathcal{U}|x_k \right] \geq 1-p_u,l\in \mathbb{N}_0\label{T6}
\end{align}
with an ambiguity set
\begin{align}
{\mathcal P}=\left\{ {\mathbb P}: {\mathbb E}_{[{\mathbb
P}]}\left[ w_{k} \right]= 0,{\mathbb E}_{\left[{\mathbb P}\right]}\left[ w_{k} w_{k}^T \right]= W \right\},\label{T4}
\end{align}
where $ {\mathcal P}$ is the set of all probability distributions on ${\mathbb P}$ and ${\mathbb E}_{[{\mathbb P}]}\left[ \cdot \right]$ denotes the expectation under the distribution ${\mathbb P}$.

Different from the existing SMPC approaches, chance constraints (\ref{T2}) and (\ref{T3}) are investigated in the DRO framework. The idea of the DRO is optimizing the `worst-case' distribution among all of the possible distributions in ${\mathcal P}$ as shown in (\ref{T5}) and (\ref{T6}).

In practical applications, the most of constraints are presented as two-sided constraints in  \cite{SHANG201924,7733074}. Therefore, it is of practical significance to study two-sided chance constraints in the context of SMPC. The two-sided DRC2s are  defined as follows
\begin{align}
&\inf_{{\mathbb P}\in{\mathcal P}} {\rm Pr}_{\left[{\mathbb P}\right]}  \left[ \left|a^T x_{l|k}\right|\leq b |x_k\right] \geq 1-p_x,\ l\in\mathbb{N}_{0};\label{T9}\\
&\inf_{{\mathbb P}\in{\mathcal P}} {\rm Pr}_{\left[{\mathbb P}\right]}  \left[ \left|c^T u_{l|k}\right|\leq d |x_k\right] \geq 1-p_u,\ l\in\mathbb{N}_{0},\label{T10}
\end{align}
with constant vectors $a \in \mathbb{R}^{n_x}$, $c \in \mathbb{R}^{n_u}$  and constants $b\in \mathbb{R}$, $d\in \mathbb{R}$. Although the expressions \eqref{T9}-\eqref{T10} are complex, we will show that they are easier to be reformulated into deterministic constraints.

\subsection{Cost Function}
The objective function is defined as a sum of quadratic stage costs, given by
\begin{align}
&J(x_k,\mathbf{u}_k)\nonumber\\
&=\mathbb{E}_{\left[{\mathbb P}\right]}\left[\sum_{l=0}^{N-1}\left(x_{l|k}^T Q x_{l|k}+ u_{l|k}^T R u_{l|k}\right)
+ x_{N|k}^TS x_{N|k}\right],\label{T11}
\end{align}
where $N\in \mathbb N$ ($\mathbb N=\{1,2,\cdots\}$) is the prediction horizon, $\mathbf{u}_k=\left[u_{0|k},u_{1|k},\cdots,u_{N-1|k}\right]$ is optimized control sequence and $ Q\in \mathbb{R}^{n_x\times n_x}$, $ R\in \mathbb{R}^{n_u\times n_u}$  are two known positive definite weighted matrices, the terminal weight matrix  $S\in \mathbb{R}^{n_x\times n_x}$ satisfies the following assumption.
\begin{assumption} The terminal weight matrix $S$ is chosen as the solution of the following Lyapunov equation
\begin{equation}\label{T12}
(A+BK)^TS(A+BK)-S=-Q-K^TRK,
\end{equation}
where $K$ is a feedback gain to be computed.
\end{assumption}

\subsection{Optimization Problem}
Now we formally state the DR-SMPC optimization problem with  two-sided DRC2s over a task horizon $N$ as {\bf DR-SMPC-P1}, which follows:
\begin{eqnarray}
\min_{{\bf u}_k} & &\!\!\!\!\!\!\!\! J(x_k,\mathbf{u}_k) \label{T13}\\
{\rm s.t.} &&\!\!\!\!\!\!\!\!\!\! x_{l+1|k}=A x_{l|k}+Bu_{l|k}+w_{l|k}, l\in\mathbb{N}_{[0,N-1]}\label{T14};\\
&&\!\!\!\!\!\!\!\!\!\! \inf_{{\mathbb P}\in{\mathcal P}} {\rm Pr}_{\left[{\mathbb P}\right]}  \left[ \left|a^T x_{l|k}\right|\leq b \right] \geq 1-p_x,   l\in\mathbb{N}_{[0,N]};\label{T15}\\
&&\!\!\!\!\!\!\!\!\!\!\inf_{{\mathbb P}\in{\mathcal P}} {\rm Pr}_{\left[{\mathbb P}\right]}  \left[ \left|c^T u_{l|k}\right|\leq d\right] \geq 1-p_u,\ l\in\mathbb{N}_{[0,N-1]};\label{T16}\\
&&\!\!\!\!\!\!\!\!\!\! x_{0|k}=x_k. \label{T166}
\end{eqnarray}
 Unfortunately, it can be observed from (\ref{T13})-(\ref{T166}) that {\bf DR-SMPC-P1}  contains several sources of intractability, that is (i) the expectation
in (\ref{T13}) is taken the unknown probability measure; (ii) the two-sided DRC2s (\ref{T15})-(\ref{T16}) are generally intractable and nonconvex. In next section, we shall develop a computational method to address these challenges, {\bf DR-SMPC-P1} is efficiently achieved by reformulating it into a conic optimization problem, which is computationally tractable.

\section{DR-SMPC Algorithm}
In this section, we shall reformulate {\bf DR-SMPC-P1} into a computationally  tractable conic optimization problem.
 To this end, we shall find a state feedback structure.
\subsection{Feedback Structure}
 We define the predicted nominal state and input as $\bar{x}_{l|k}$, $\bar{u}_{l|k}$, and the nominal dynamics model of (\ref{YT1}) is expressed as
\begin{eqnarray}\label{T17}
\bar{x}_{l+1|k}=A\bar{x}_{l|k}+B\bar{u}_{l|k}.
\end{eqnarray}
 To obtain the computable form of cost function and two-sided DRC2s, a state feedback control law \citep{2005Robust,HEWING2020109095} is designed as following
\begin{eqnarray}\label{T18}
u_{l|k}=K(x_{l|k}-\bar{x}_{l|k})+\bar{u}_{l|k},
\end{eqnarray}
where $K$ is a selected feedback gain, the term $\bar{u}_{l|k}$ replaces $u_{l|k}$ as the new decision variable in {\bf DR-SMPC-P1}.

A stochastic error between the real state $x_k$ and the nominal state $\bar{x}_k$ at time $k$ is commonly denoted as $\Delta x_k=x_k-\bar{x}_k$.
Based on (\ref{YT1}), (\ref{T17}) and (\ref{T18}),  the state error dynamic system is expressed as:
\begin{eqnarray}
\Delta x_{l+1|k}=(A+BK)\Delta x_{l|k}+w_{l|k}.\label{T20}
\end{eqnarray}
At time step $k=0$, a proper initialization is used, i.e., $\bar{x}_0=x_0$, and thus recalling that the noise
is zero mean, the expected value of the stochastic error is $\mathbb{E}_{\left[{\mathbb P}\right]}[\Delta x_{k}]=0$, thus, the predicted covariance matrix $\Sigma_{l|k}$ is able to be updated as:
\begin{eqnarray}\label{T21}
\Sigma_{l+1|k}&=&\mathbb{E}_{\left[{\mathbb P}\right]}\left[\Delta x_{l+1|k} \Delta x_{l+1|k}^T\right]\nonumber\\
&=&(A+BK)\Sigma_{l|k}(A+BK)^T+W.
\end{eqnarray}

The equivalent sets of the two-sided DRC2s are formed by simply substituting $x_{l|k} =\bar{x}_{l|k}+\Delta x_{l|k}$ and $u_{l|k}=\bar{u}_{l|k}+ K\Delta x_{l|k}$ into \eqref{T15}-\eqref{T16}, which gives
\begin{align}
&\!\!\!\!\!\!\bar{X}_{l|k}=\left\{\bar{x}_{l|k}\big|\inf_{{\mathbb P}\in{\mathcal P}} {\rm Pr}_{\left[{\mathbb P}\right]}  \left[ \left|a^T(\bar{x}_{l|k}+\Delta x_{l|k})\right|\leq b\right] \geq 1-p_x\right\}\nonumber\\
&\ \ \ \  \ \ \ \ \  \ \ \  \ \ \ \ \ \  \ \ \ \  \ \ \ \ \  \ \ \  \ \ \ \ \ \ \ \ \ \  \ \ \ \ \  \ \ \   l\in\mathbb{N}_{[0,N]};\label{T22}\\
&\!\!\!\!\!\!\bar{U}_{l|k}=\left\{\bar{u}_{l|k}\big| \inf_{{\mathbb P}\in{\mathcal P}} {\rm Pr}_{\left[{\mathbb P}\right]}  \left[ \left|c^T(\bar{u}_{l|k}+ K\Delta x_{l|k})\right|\leq d \right] \geq 1-p_u\right\}\nonumber\\
&\ \ \ \  \ \ \ \ \  \ \ \  \ \ \ \ \ \  \ \ \ \  \ \ \ \ \  \ \ \  \ \ \ \ \ \ \ \ \ \  \ \ \ \ \ \ \ \  l\in\mathbb{N}_{[0,N-1]}.\label{T23}
\end{align}

\subsection{ Convex Formulation of  DR-SMPC-P1}
Based on the results above, we show that {\bf DR-SMPC-P1} can be represented as a  convex cone program problem and
hence is computationally tractable.
\newtheorem{theorem}{Theorem}
\begin{theorem}
{\bf DR-SMPC-P1} is exactly reformulated as following conic optimization problem, which are referred to as {\bf DR-SMPC}
\begin{eqnarray}
\min_{\bar{{\bf u}}_k,y_{1,l},\lambda_{1,l},y_{2,l},\lambda_{2,l}}&&  \!\!\!\!\!\!\!\!\!\!\!\!\sum_{l=0}^{N-1}\left[\bar{x}_{l|k}^T Q \bar{x}_{l|k} +\bar{u}_{l|k}^T R \bar{u}_{l|k} \right]+\bar{x}_{N|k}^TS \bar{x}_{N|k}\nonumber\\
&&\!\!\!\!\!\!\!\!\!\!\!\!\!\!\!\!\!\!\!\!\!\!\!\!\!\!\!\!\!\!\!\!\!\!\!\!\!\!\!\!\!\!+ \sum_{l=0}^{N-1}\text{trace}\left[\left(Q+K^TRK\right)\Sigma_{l|k}\right]+\text{trace}\left(S\Sigma_{N|k}\right)\label{W26}\\
&& \!\!\!\!\!\!\!\!\!\!\!\!\!\!\!\!\!\!\!\!\!\!\!\!\!\!\!\!\!\!\!\!\!\!\!\!\!\!\!\!\!\!\!\!\!\!\!\!\!\!\!\!{\rm s.t.}\ \ \ \bar{x}_{l+1|k}=A \bar{x}_{l|k}+B\bar{u}_{l|k}, l\in\mathbb{N}_{[0,N-1]},\\
&&\!\!\!\!\!\! \!\!\!\!\!\!\!\!\!\!\!\!\!\!\!\!\!\!\!\!\!\!\!\!\!\!\!\!\!\!\!x_{l+1|k}=A x_{l|k}+Bu_{l|k}+w_{l|k}, l\in\mathbb{N}_{[0,N-1]},\\
&&\!\!\!\!\!\!\!\!\!\!\!\!\!\!\!\!\!\!\!\!\!\!\!\!\!\!\!\!\!\!\!\!\!\!\!\!\!u_{l|k}=K (x_{l|k}-\bar{x}_{l|k})+\bar{u}_{l|k},l\in\mathbb{N}_{[0,N-1]},\\
&&\!\!\!\!\!\!\!\!\!\!\!\!\!\!\!\!\!\!\!\!\!\!\!\!\!\!\!\!\!\!\!\!\!\!\!\!\!\bar{x}_{l|k}\in \bar{X}_{l|k},l\in\mathbb{N}_{[0,N-1]},\\
&&\!\!\!\!\!\!\!\!\!\!\!\!\!\!\!\!\!\!\!\!\!\!\!\!\!\!\!\!\!\!\!\!\!\!\!\!\!\bar{u}_{l|k}\in \bar{U}_{l|k},l\in\mathbb{N}_{[0,N-1]},\\
&&\!\!\!\!\!\!\!\!\!\!\!\!\!\!\!\!\!\!\!\!\!\!\!\!\!\!\!\!\!\!\!\!\!\!\!\!\!\bar{x}_{N|k}\in \bar{\mathcal{X}}_f.\label{W27}
\end{eqnarray}
where $\bar{{\bf u}}_k=\left[\bar{u}_{0|k},\bar{u}_{1|k},\cdots,\bar{u}_{N-1|k}\right]$, $y_{(1,l)}, y_{(2,l)}, \lambda_{(1,l)}$  and $\lambda_{(2,l)}$ are the newly introduced decision variable, $\bar{\mathcal{X}}_f$ is terminal constraint to guarantee recursive feasibility and  convergence  of the proposed approach,
\begin{align}\label{TY10}
\bar{X}_{l|k}:=\left\{\bar{x}_{l|k}\Bigg|\begin{aligned}
&y_{(1,l)}^2+a^T\Sigma_{l|k} a \leq p_x(b-\lambda_{(1,l)})^2,\\
&\left|a^T\bar{x}_{l|k}\right|\leq y_{(1,l)}+\lambda_{(1,l)},\\
& y_{(1,l)}\geq 0, 0\leq\lambda_{(1,l)} \leq b
\end{aligned}\right\}\nonumber
\end{align}
and
\begin{align}\label{TY20}
\bar{U}_{l|k}:=\left\{\bar{u}_{l|k}\Bigg|\begin{aligned}
&
y_{(2,l)}^2+c^TK^T\Sigma_{l|k}K c \leq p_u(d-\lambda_{(2,l)})^2,\\
&\left|c^T\bar{u}_{l|k}\right|\leq y_{(2,l)}+\lambda_{(2,l)}, \\
&y_{(2,l)}\geq 0,0\leq\lambda_{(2,l)} \leq d
\end{aligned}\right\}.\nonumber
\end{align}
\end{theorem}
{\sc Proof} See Appendix A.

Note that {\bf DR-SMPC} is now converted to a conic optimization problem, which is computationally tractable and can be solved by using standard software package such as CVX \citep{2004Convex}. The optimization is carried out over the optimal nominal control sequence $\bar{\mathbf{u}}_k^*$, the first element $\bar{u}_k^*$ of $\bar{\mathbf{u}}_k^*$ is applied to the the feedback control law $u_k^*=K(x_k-\bar{x}_k)+\bar{u}_k^*$. This is crucial to solve the two-sided chance constrained control problem. However, the determination of the terminal set as well as the feasibility set of nominal initial states is rather difficult but particularly important for establishing recursive feasibility and convergence of the proposed approach.
\subsection{Determination of Terminal Constraint}
It is well-known that terminal constraint is closely related to the  stability of the nominal system. Thus, the following terminal constraint is imposed
 \begin{equation}\label{TTT10}	
\bar{x}_{N|k}\in \bar{\mathcal{X}}_f\subseteq \bar{X}_{l|k},
 \end{equation}
and the set $\bar{\mathcal{X}}_f$ is a positively invariant set satisfying
\begin{align}\label{TTT102}
\bar{x}_{N+1|k}=(A+BK)\bar{x}_{N|k}\in \bar{\mathcal{X}}_f, \forall \bar{x}_{N|k}\in \bar{\mathcal{X}}_f.
\end{align}

To achieve the recursive feasibility of the proposed approach in Section IV, the terminal set is chosen as following
\begin{eqnarray}
\bar{\mathcal{X}}_f=\left\{\bar{x}_{N|k}\Bigg|\begin{aligned}
&
y^2+a^T\bar{\Sigma}a \leq p_x(b-\lambda)^2,0\leq\lambda \leq b,\\
&\left|a^T\bar{x}_{N|k}\right|\leq y+\lambda, y\geq 0
\end{aligned}\right\},
\end{eqnarray}
where $y, \lambda$ are extra variables, $\bar{\Sigma}$ is the steady state solution of the Lyapunov equation (\ref{T21}) computed as following
\begin{eqnarray}\label{TTT101}
\bar{\Sigma}=(A+BK)\bar{\Sigma}(A+BK)^T+W.
\end{eqnarray}

To facilitate the proof of recursive feasibility,  we explicitly point out the technical assumption that is necessary.

\begin{assumption} The selected feedback gain $K$ satisfies\\
(a). $K\odot \bar{\mathcal{X}}_f \in \bar{U}_{N|k}$;\\
(b). $|\lambda_i(A+BK)| < 1$, where $\lambda_{i}(A+BK)$ is $i$th eigenvalue of $A+BK$.
\end{assumption}

It can be observed from (\ref{T21}) and (\ref{TTT101}) that $\Sigma_{N|k}\leq \bar{\Sigma}$, let $\lambda_{(1,l)}=\lambda$, then, there must be $0\leq y\leq y_{(1,l)}$ such that $y_{(1,l)}^2+a^T\Sigma_{l|k} a \leq p_x(b-\lambda_{(1,l)})^2$ is equivalent to $y^2+a^T\bar{\Sigma} a \leq p_x(b-\lambda)^2$, therefore,
it must hold $\bar{\mathcal{X}}_f \subseteq \bar{X}_{l|k}$. According to Assumption 2, there must be $\bar{u}_{N|k}=K\bar{x}_{N|k}\in \bar{U}_{N|k}$ such that $\left|a^T(A+BK)\bar{x}_{N|k}\right|\leq \left|a^T\bar{x}_{N|k}\right|\leq y+\lambda$. The chosen terminal set $\bar{\mathcal{X}}_f$ satisfies the conditions of (\ref{TTT10}) and (\ref{TTT102}).

\subsection{Binary Initialization Strategy}
It is quite clear that the initial condition $\bar{x}_k$ is critical to performance index. At each step, the most recent information available on the real state should be used to reset the nominal state $\bar{x}_k$. Specifically, the selected ``optimal'' current value of $\bar{x}_k$ is set to $x_k$. However, the possibility of unbounded noises cannot be completely ruled out, the choice of $\bar{x}_k=x_k$ may cause an infeasible optimization problem, and the basic property of recursive feasibility would be lost. Therefore, the following binary initialization strategy is defined to
guarantee the recursive feasibility.

{\it Strategy 1} - Using the most recent information available on the measured state at time step $k$, i.e., $\bar{x}_k=x_k$.

{\it Strategy 2} - Using the updating information according to the following past optimal solution, i.e., $\bar{x}_k=\bar{x}_{k|k-1}$,
\begin{align}
\bar{x}_{k|k-1}=A\bar{x}_{k-1|k-1}+B\bar{u}_{k-1|k-1}.
\end{align}

The choice of binary initialization strategy requires to solve two optimization problems at each time step, i.e., {\bf DR-SMPC} with {\it Strategy 1} and {\bf DR-SMPC} with {\it Strategy 2}. The rules for adopting {\it Strategy 1} and {\it Strategy 2} are stated as follows: {\bf DR-SMPC} with {\it Strategy 1} is first solved, then, if it is infeasible, {\bf DR-SMPC} with {\it Strategy 2} must be executed. On the contrary, if it is feasible, the optimal cost between {\bf DR-SMPC} with {\it Strategy 1} and {\it Strategy 2} need to be compared. If the optimal cost of {\it Strategy 1} is higher, {\bf DR-SMPC} with {\it Strategy 2} should be solved. Although the binary initialization strategy does not guarantee  optimality, the convergence properties of the proposed method is obtained.

\subsection{DR-SMPC Algorithm}
Based on Theorem 1 and binary initialization strategy of nominal state, the implementation of the DR-SMPC algorithm is summarized
as Algorithm 1 given below.
\begin{algorithm}[htb]
 \caption{DR-SMPC}
 \begin{algorithmic}[1] %这个1 表示每一行都显示数字
  \REQUIRE ~~\\ %算法的输入参数：Input
  {\bf Input}: $A$, $B$, $a$, $b$, $c$, $d$, $W$, $Q$, $R$ $x_k$, $\bar{x}_k$, $p_x$,  $p_u$ and $N$;\\
  {\bf Output}: $x_{k+1}$, $\bar{x}_{k+1}$;\\
  {\bf Initialize}: $\bar{x}_k=x_k$ at time step $k=0$, {\bf DR-SMPC} is feasible;\\
  {\bf Off-line}: Compute the feedback gain $K$ and the terminal weight $S$ according to assumption 1 and assumption 2;\\
   {\bf On-line}:
  \WHILE{1}
  \IF{{\bf DR-SMPC} with {\it Strategy 1} is infeasible}
  \STATE {\bf DR-SMPC} with {\it Strategy 2} is executed;\
  \ELSE
  \STATE Compare the optimal cost between {\bf DR-SMPC} with {\it Strategy 1} and {\bf DR-SMPC} with {\it Strategy 2};\
  \IF{the optimal cost of {\bf DR-SMPC} with {\it Strategy 1} is higher}
  \STATE {\bf DR-SMPC} with {\it Strategy 2} is executed;\
  \ENDIF
  \ENDIF
  \STATE Acquire nominal input $\bar{u}_k^*$ and the control input $u_k^*=K(x_k-\bar{x}_k)+\bar{u}_k^*$;\
  \STATE Compute $\bar{x}_{k+1}$ according to (7), and measure the state $x_{k+1}$;
  \STATE Update $k=k+1$;
  \ENDWHILE
 \end{algorithmic}
\end{algorithm}

\section{Recursive Feasibility and Convergence}
As an important property of the SMPC, the recursive feasibility is a precondition for the convergence properties of the stochastic dynamic system, which makes great impacts on the implementation of controller.
\subsection{Recursive Feasibility}
The following theorem plays an important role to establish recursive feasibility of the proposed DR-SMPC algorithm.
\begin{theorem}If {\bf DR-SMPC} with {\it Strategy 2} is feasible at time step $k$, then, {\bf DR-SMPC} with {\it Strategy 2} remains feasible at each time step.
\end{theorem}
{\sc Proof} See Appendix B.

\subsection{Convergence}
Since the optimal cost of {\bf DR-SMPC} with {\it Strategy 2} is lower, {\bf DR-SMPC} with {\it Strategy 2} is uesed. Therefore,
we state the main result concerning the convergence properties of the algorithm by the following theorem.
\begin{theorem}  Let $J(\bar{x}_k,\bar{\mathbf{u}}_k^*)$ be the optimal cost of {\bf DR-SMPC} with {\it Strategy 2} at time step $k$, then
\begin{eqnarray}
&&\!\!\!\!\!\!\!\!\!\!J(\bar{x}_{k+1},\bar{\mathbf{u}}_{k+1}^*)-J(\bar{x}_{k},\bar{\mathbf{u}}_{k}^*)\nonumber\\
&&\leq -\mathbb{E}_{\left[{\mathbb P}\right]}\left[x_k^TQx_k+u_k^TRu_k\right]+\text{trace}(SW).
\end{eqnarray}
\end{theorem}
{\sc Proof} See Appendix C.

According to theorem 3 and similar arguments to the paper \citep{7733074,TY}, we can give a standard argument following as
\begin{eqnarray}
\lim_{N \rightarrow \infty} \frac{1}{N}\sum_{k=0}^N \mathbb{E}_{\left[{\mathbb P}\right]}\left[x_k^TQx_k+u_k^TRu_k\right]\leq \text{trace}(SW)
\end{eqnarray}
This indicates that the state of the system is driven to a neighborhood of the steady state condition. Then, the {\bf DR-SMPC} algorithm is convergent for system (\ref{T1}) under the control law (\ref{T18}).

\section{Simulation Example}
In this section, two practical control systems under different types of noises are provided to tested the performance of the proposed DR-SMPC algorithm. Specifically, the noise is subject to Gaussian distribution in a Buck-Boost DC-DC power converters. To further show the superior of the proposed algorithm, we consider a two-mass spring system in which the noise is subject to Laplace distribution in the second example. The proposed DR-SMPC algorithm is compared with G-SMPC algorithm  in \cite{2012Nonquadratic} and P-SMPC algorithm  in \cite{2013Ao}.

\subsection{A Buck-Boost DC-DC Power Converters}
 In this example, an ideal Buck-Boost circuit is considered \citep{7733074}, the state $x_k=[x_k^1,x_k^2]^T$, and the corresponding matrices are
 \begin{align*}
A=
\left[\begin{matrix}
1 & 0.0075\\
-0.143  & 0.996
\end{matrix}\right],\
B=
\left[\begin{matrix}
4.798\\
0.115
\end{matrix}\right].
\end{align*}

It is well known that the current through the inductor violates the rated current with a small probability will not cause damage to its life cycle. Similarly, it is also practical to express the voltage constraint of resistance as chance constraint. The chance constraints on state and input are given as
\begin{align}
&{\rm Pr}_{\left[{\mathbb P}\right]}\left\{| x_k^1|\leq2\right\}\geq 0.8,\\
&{\rm Pr}_{\left[{\mathbb P}\right]}\left\{| x_k^2|\leq3\right\}\geq 0.8,\\
 &{\rm Pr}_{\left[{\mathbb P}\right]}\left\{|u_k|\leq 0.2\right\}\geq 0.99.
\end{align}

The control objective of this example is to design a MPC control law to regulate the state to near the origin in the presence of  the noise.  The prediction horizon is chosen as $N=8$, ${w}_k \sim {\mathcal {N}}(0,0.03I_2)$, $Q= \text{diag}\{1,10\}$, $R= 1$, $S=\left[\begin{matrix}
1.90& -5.05\\
-5.05& 39.54
\end{matrix}\right]$ and the selected $K=[-0.28,	0.49]$.
\begin{figure}
\centering
\includegraphics*[width=7cm]{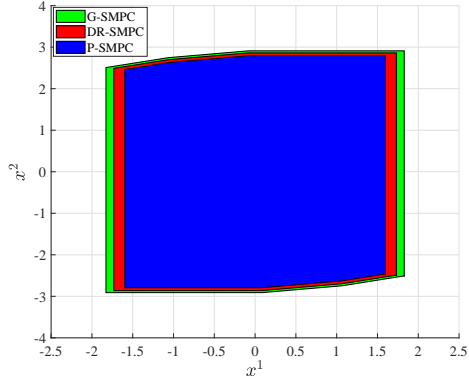}
\centering
\caption{Plots of the feasibility sets for G-SMPC, DR-SMPC and P-SMPC algorithm  in example 1.}\label{fig222}
\end{figure}
In order to state the fact that feasibility set of the proposed algorithm increases, we define the feasibility set of the initial nominal state on SMPC algorithm as following
\begin{align}
F=\{ \bar{x}_0\in \mathcal{X} | \text{SMPC with  initial  states} \ \bar{x}_0 \ \text{is feasible} \}.\nonumber
\end{align}
In Fig. \ref{fig222}, we compare the feasible sets obtained with the proposed DR-SMPC algorithm,  G-SMPC algorithm and P-SMPC algorithm.
Apparently, the feasible set of the proposed DR-SMPC algorithm has 1.15 times the size of the feasible set of P-SMPC algorithm, in view of the reformulation of two-sided chance constraints, the P-SMPC algorithm is more conservative than the DR-SMPC algorithm. On the other hand, the feasible set of the G-SMPC algorithm is larger than these of the the DR-SMPC and P-SMPC algorithms since the G-SMPC algorithm utilizes the distribution information of the noise.

\subsection{A two-mass spring system}
\begin{figure}
\centering
\includegraphics*[width=7cm]{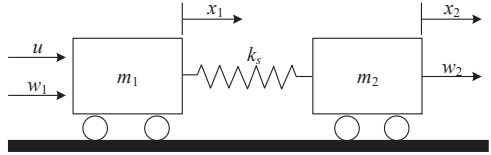}
\centering
\caption{Two-mass spring system.}\label{fig22}
\end{figure}
A single spring and double mass system \citep{SHANG201924} is illustrated in Fig. \ref{fig22}. In this figure, the two blocks with mass $m_1$ and $m_2$ are linked by springs of coefficient $k_s$. The manipulated variable is the applied force $u$ and the output variables are the distances and speed of the mass block movement $x^1$, $x^3$ corresponding to block one and $x^2$, $x^4$ corresponding to block two. $w^1$ and $w^2$ are external uncertainties acting on $m_1$ and $m_2$ , respectively. By applying Newton's law to the first and second block, we obtain
 \begin{align}
&\dot{x}_t^1=x_t^3,\\
&\dot{x}_t^2=x_t^4,\\
&m_1\dot{x}_t^3=-k_s(x_t^1-x_t^2)+u_t+w_t^1,\\
&m_2\dot{x}_t^4=k_s(x_t^1-x_t^2)+w_t^2.
\end{align}
Defining the state as  $x = [x^1, x^2, x^3, x^4]$, $w = [w^1, w^2]$,  the discrete state space by Euler's approximation method for a sampling time $T_s$ is
 \begin{align}
x_{k+1}=Ax_k+Bu_k+w_k.
\end{align}
where the two blocks mass $m_1=1$, $m_2=1$,  the elastic constant $k_s=1.25$ and sampling time $T_s=0.1$.  In each step, $w_k$  is subject to Laplace distribution
with zero-mean and variance $0.07I_2$. The quadratic cost matrices $Q=\text{diag}(1,1,4,6)$, $R=1$ reflect the expensiveness of thruster use in space. The initial condition is chosen as $ x_0= [0.5,0.5,0,0]^T$ towards the origin as the desired state, the prediction horizon $N=7$.  For keeping the state steady, we consider constraints on state and input as
\begin{align}
&{\rm Pr}_{\left[{\mathbb P}\right]}\left\{| x_k^3|\leq0.12\right\}\geq 0.8,\\
&{\rm Pr}_{\left[{\mathbb P}\right]}\left\{| x_k^4|\leq0.12\right\}\geq 0.8,\\
 &{\rm Pr}_{\left[{\mathbb P}\right]}\left\{|u_k|\leq 0.5\right\}\geq 0.99.
\end{align}

\begin{figure}
\centering
\includegraphics*[width=7cm]{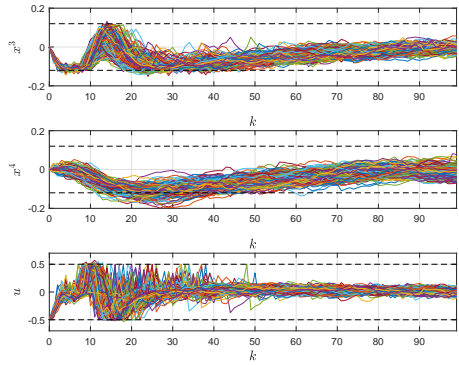}
\centering
\caption{The trajectories of G-SMPC algorithm in example 2.}\label{fig1:side:a}
\end{figure}
\begin{figure}
\centering
\includegraphics*[width=7cm]{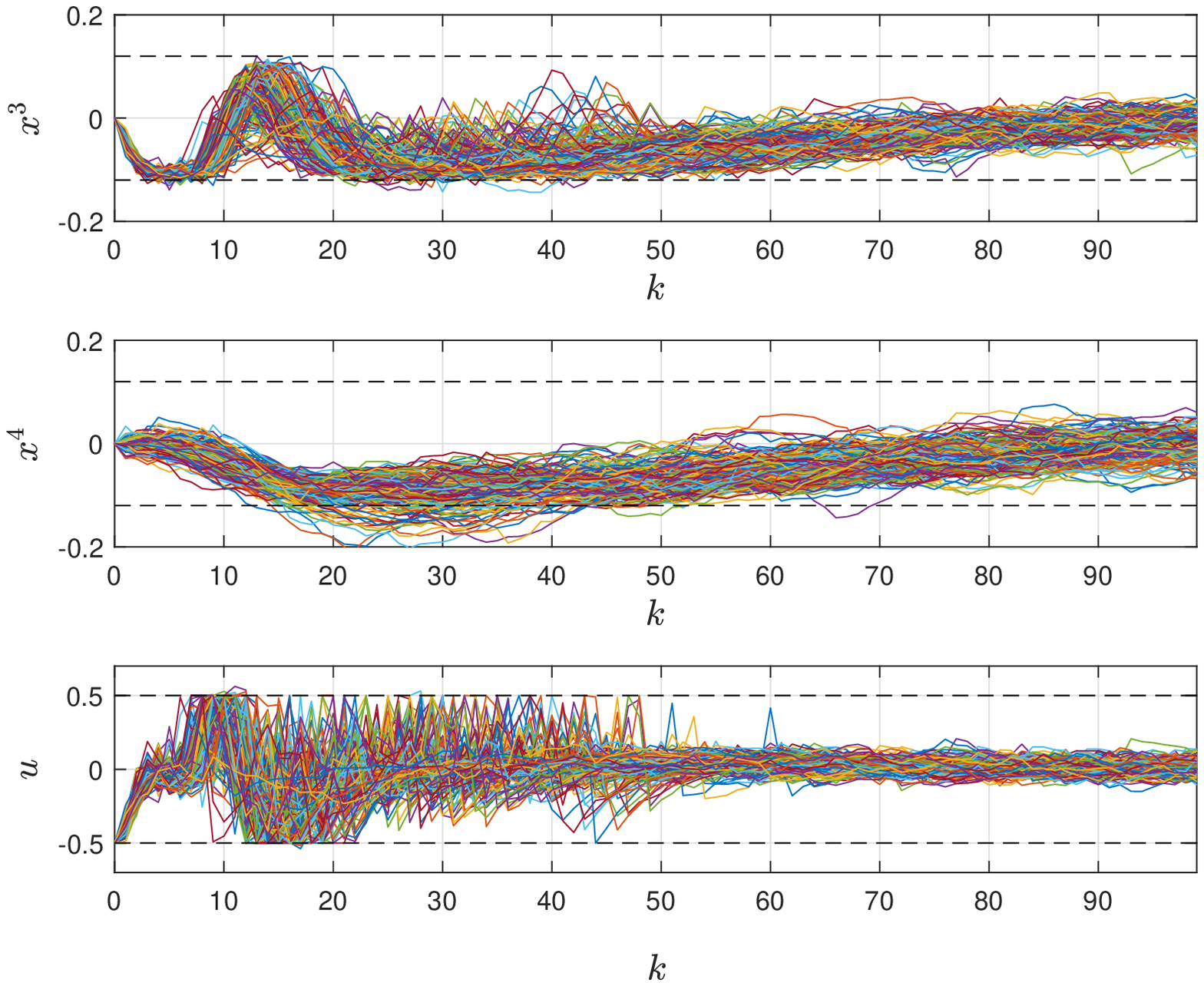}
\centering
\caption{The trajectories of DR-SMPC algorithm in example 2.}\label{fig1:side:b}
\end{figure}
\begin{figure}
\centering
\includegraphics*[width=7cm]{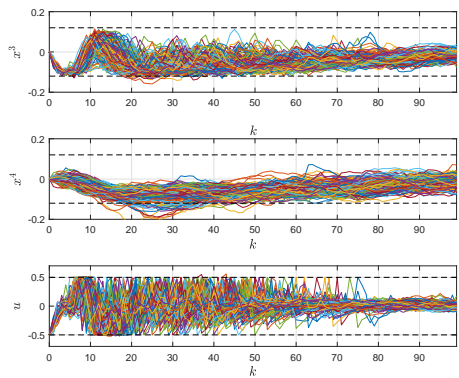}
\centering
\caption{The trajectories of P-SMPC algorithm in example 2.}\label{fig1:side:c}
\end{figure}
\begin{figure}
\centering
\includegraphics*[width=7cm]{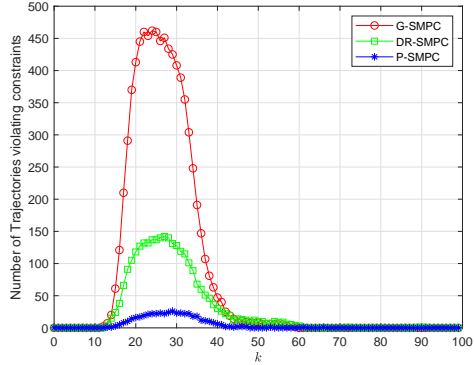}
\centering
\caption{Number of trajectories violating constraints in each time step in example 2.}\label{fig1:side:d}
\end{figure}

The G-MPC algorithm, P-SMPC algorithm with fixed uniform risk allocation and the proposed DR-SMPC algorithm based on a Monte-Carlo simulation of 1000 runs in Fig. \ref{fig1:side:a}-\ref{fig1:side:c}. The result of is violating constraints is provided in Fig. \ref{fig1:side:d}, in out of 1000 sample trajectories,  the biggest number of sample trajectories violating the constraints for the three algorithms are respectively 462, 149 and 28. The probability violating the constraints on the P-SMPC algorithm is  $2.8\%$, this is due to the two-sided chance constraints is approximately broken into two single chance constraints that lead to conservatism
However, the violation probabilities of G-MPC algorithm exceeds the set value $20\%$, since deviation of the assumed distribution from the true one caused by poor assumptions result in unwanted behavior of the system. As a consequence, we can draw a conclusion that P-SMPC algorithm is more conservative than DR-SMPC in the reformulation of two-sided chance constraints. For the G-SMPC algorithm, the assumption of gaussian noise distribution may not be consistent with the real scene, which will lead to the opposite result.
\section{Conclusions}
This paper advocates a DRO approach to SMPC under unbounded stochastic nosie following an unknown
distribution. A conic representation of two-sided DRC2s is obtained, which contributes to a
tractable formulation of the SMPC problem, based on which the recursive
feasibility and convergence  are established. Furthermore, numerical results showed that the performance of the proposed algorithm is much better than the P-SMPC algorithm with fixed uniform risk allocation and G-SMPC.

\section*{Appendix}
\subsection{Proof of Theorem 1}
Note that $x_{l|k}=\Delta x_{l|k}+\bar{x}_{l|k}$, $u_{l|k}=\bar{u}_{l|k}+K\Delta x_{l|k}$, $\mathbb{E}_{\left[{\mathbb P}\right]}[\Delta x_{l|k}]=0$  in (\ref{T20}), $\mathbb{E}_{\left[{\mathbb P}\right]}[\Delta x_{l|k}^TQ\Delta x_{l|k}]=\text{trace}(Q\Sigma_{l|k})$, $\mathbb{E}_{\left[{\mathbb P}\right]}[\Delta x_{N|k}^TS\Delta x_{N|k}]=\text{trace}(S\Sigma_{N|k})$. As consequence, Eq.\eqref{T11} is equivalent to the following equation
\begin{align}
&J(\bar{x}_k,\bar{\mathbf{u}}_k)\nonumber\\
& = \sum_{l=0}^{N-1}\left[\bar{x}_{l|k}^T Q \bar{x}_{l|k} +\bar{u}_{l|k}^T R \bar{u}_{l|k} \right]+\bar{x}_{N|k}^TS \bar{x}_{N|k}\nonumber\\
&\ \ \ \ +\  \sum_{l=0}^{N-1}\text{trace}\left[\left(Q+K^TRK\right)\Sigma_{l|k}\right]+\text{trace}\left(S\Sigma_{N|k}\right). \label{T26}
\end{align}

We shall use DRO approach to reformulate the two-sided chance constraints  into SOC form. Moreover, we shall show that this reformulation is exact according to  following lemma \citep{2016Distributionally}.
\newtheorem{lemma}{Lemma}
\begin{lemma}
The two-sided DRC2s
\begin{align}
\inf_{{\mathbb P}\in{\mathcal P}} {\rm Pr}_{\left[{\mathbb P}\right]} \left[\left|a(x)^Tw+b(x)\right|\leq T\right]\geq 1-\epsilon
\end{align}
is reformulated into the following convex SOC constraints
\begin{align}
&y^2+a(x)^T\Sigma a(x) \leq \epsilon (T-\lambda)^2,\\
& |b(x)|\leq y+\lambda,\\
&T\geq \lambda\geq0, y\geq0.
\end{align}
where $a(x)$ and $b(x)$ are function of decision variable $x$, and $y, \lambda$ are involving two additional variables, uncertainty $w$ is  satisfying
\begin{align}
&\!\!\!\!{\mathcal P}=\left\{ {\mathbb P}:  \mathbb{E}_{\left[{\mathbb P}\right]}\left[w\right]=0,\ \mathbb{E}_{\left[{\mathbb P}\right]}\left[ww^T\right]=\Sigma \right\}.
\end{align}
\end{lemma}

Let $a(x)=a, b(x)=a^T\bar{x}_{l|k}, w=\Delta x_{l|k}$ and $T=b$ according to Lemma 1,  \eqref{T22} is  reformulated into the following
\begin{eqnarray}
\bar{X}_{l|k}:=\left\{\bar{x}_{l|k}\Bigg|\begin{aligned}
&
y_{(1,l)}^2+a^T\Sigma_{l|k} a \leq p_x(b-\lambda_{(1,l)})^2,\\
&\left|a^T\bar{x}_{l|k}\right|\leq y_{(1,l)}+\lambda_{(1,l)}, \\
&y_{(1,l)}\geq 0,0\leq\lambda_{(1,l)} \leq b
\end{aligned}\right\}.
\end{eqnarray}
In a similar manner, \eqref{T23} is  reformulated into the following
\begin{eqnarray}\label{TY200}
\bar{U}_{l|k}:=\left\{\bar{u}_{l|k}\Bigg|\begin{aligned}
&
y_{(2,l)}^2+c^TK^T\Sigma_{l|k}K c \leq p_u(d-\lambda_{(2,l)})^2,\\
&\left|c^T\bar{u}_{l|k}\right|\leq y_{(2,l)}+\lambda_{(2,l)}, \\
&y_{(2,l)}\geq 0,0\leq\lambda_{(2,l)} \leq d
\end{aligned}\right\}.
\end{eqnarray}
This completes the proof.

\subsection{Proof of Theorem 2}
 At time step $k$, let  {\bf DR-SMPC} with {\it Strategy 2} is feasible, the corresponding optimal solution
 is the nominal control sequence $\bar{\mathbf{u}}_{k}^*=[\bar{u}_{0|k}^*,\bar{u}_{1|k}^*,\cdots,\bar{u}_{N-1|k}^*]$ with the current and predicted nominal state trajectory  $\bar{\mathbf{x}}_{k}^*=[\bar{x}_{0|k},\bar{x}_{1|k}^*,\bar{x}_{2|k}^*,$ $\cdots,\bar{x}_{N-1|k}^*,\bar{x}_{N|k}^*]$.

At next time step $k+1$, the sub-optimal control sequence and the corresponding state trajectory are denoted by $\bar{\mathbf{u}}_{k+1}=[\bar{u}_{1|k}^*,\bar{u}_{2|k}^*,\cdots,\bar{u}_{N-1|k}^*, K\bar{x}_{N|k}^*]$ and $[\bar{x}_{1|k}^*,\bar{x}_{2|k}^*,\cdots,\bar{x}_{N|k}^*,$ $\bar{x}_{N+1|k}]$. According to (a) in Assumptions 2, $K\bar{x}_{N|k}^*\in \bar{U}_{N|k}$ for $\forall \ x_{N|k}^*\in \bar{\mathcal{X}}_f$, there must be $$\bar{x}_{N|k+1}=\bar{x}_{N+1|k}=(A+BK)x_{N|k}^*\in \bar{\mathcal{X}}_f.$$

\subsection{Proof of Theorem 3} At time $k+1$,  the suboptimal solution is denoted as
$$\bar{\mathbf{u}}_{k+1}=[\bar{u}_{1|k}^*,\bar{u}_{2|k}^*,\cdots,\bar{u}_{N-1|k}^*, K\bar{x}_{N|k}^*].$$

The optimal cost function value at $k+1$ is denoted as $J\left(\bar{x}_{k+1},{\bar{\bf u}}_{k+1}^*\right)$.
Considering the optimality of $J$, we have
\begin{align}\label{TY1}
J\left(\bar{x}_{k+1},\bar{{\bf u}}_{k+1}^*\right)\leq J\left(\bar{x}_{k+1},\bar{{\bf u}}_{k+1}\right).
\end{align}
According to \eqref{T26}, we have
\begin{align}
&\ \ \ \ J\left(\bar{x}_{k+1},\bar{{\bf u}}_{k+1}\right)\nonumber\\
&=J\left(\bar{x}_{k},\bar{{\bf u}}_{k}^*\right)-\bar{x}_k^TQ\bar{x}_k-\bar{u}_k^TR\bar{u}_k-\bar{x}_{N|k}^TS\bar{x}_{N|k}\nonumber\\
&\ \ \ \ -\text{trace}\left[\left(Q+K^TRK\right)\Sigma_{k}\right]-\text{trace}\left(S\Sigma_{N|k}\right)\nonumber\\
&\ \ \ \ +\bar{x}_{N|k}^TQ\bar{x}_{N|k}+\bar{x}_{N+1|k}^TS\bar{x}_{N+1|k}+\bar{u}_{N|k}^TR\bar{u}_{N|k}\nonumber\\
&\ \ \ \ +\text{trace}\left[\left(Q+K^TRK\right)\Sigma_{N|k}\right]+\text{trace}\left(S\Sigma_{N+1|k}\right).
\end{align}
In view of the definition of $S$, given by (\ref{T12}), we have also
\begin{align}
&\bar{x}_{N|k}^TQ\bar{x}_{N|k} -\bar{x}_{N|k}^TS\bar{x}_{N|k}+\bar{x}_{N+1|k}^TS\bar{x}_{N+1|k}+\bar{u}_{N|k}^TR\bar{u}_{N|k}\nonumber\\
&=\bar{x}_{N|k}^T[(A+BK)^TS(A+BK)+Q+K^TRK-S)\bar{x}_{N|k}\nonumber\\
&=0,
\end{align}
and
\begin{align}
& \text{trace}\left(S\Sigma_{N+1|k}\right)-\text{trace}\left(S\Sigma_{N|k}\right)\nonumber\\
& +\text{trace}\left[\left(Q+K^TRK\right)\Sigma_{N|k}\right]=\text{trace}\left(SW\right).
\end{align}
 Therefore, we have that
\begin{eqnarray}
&&\!\!\!\!\!\!\!\!\!\!J(\bar{x}_{k+1},\bar{\mathbf{u}}_{k+1}^*)-J(\bar{x}_{k},\bar{\mathbf{u}}_{k}^*)\nonumber\\
&&\leq -\mathbb{E}_{\left[{\mathbb P}\right]}\left[x_k^TQx_k+u_k^TRu_k\right]+\text{trace}(SW).
\end{eqnarray}

\end{document}